# Network Structures of Collective Intelligence:
# The Contingent Benefits of Group Discussion

Joshua Becker[a,1], Abdullah Almaatouq[b], and Emőke-Ágnes Horvát[c]

[a]UCL School of Management, University College London
[b]Sloan School of Management, Massachusetts Institute of Technology
[c]School of Communication, Northwestern University

[1]Correspondence to: joshua.becker@ucl.ac.uk

WORKING PAPER: last updated 5 March 2021

**Abstract.** A central interest for researchers studying numeric estimation such as forecasting is whether crowdsourced estimates are more accurate when contributors are independent or when they can influence each other. However, experimental research has produced contradictory findings on whether and when communication between contributors will improve the accuracy of the average estimate. While some evidence suggests that controlled processes where people exchange only the numeric value of their estimates ("numeric exchange") can improve accuracy more than informal discussion, others argue that discussion outperforms numeric exchange. Still others argue that independent individuals produce the most accurate numeric estimates. The present paper shows how network theories of opinion formation can resolve these contradictions, even when groups lack apparent structure as in informal discussion. Emergent network structures of influence interact with the pre-communication estimate distribution to moderate the effect of communication on accuracy. Discussion generates centralization in this emergent network which sometimes increases and sometimes decreases accuracy depending on the pre-communication estimate distribution. In contrast, numeric exchange has a relatively stable effect on accuracy. As a result, whether discussion outperforms numeric exchange depends on the statistical properties of pre-communication estimate. These results resolve contradictions in previous research and offer practical recommendations for teams and organizations.

## 1. Introduction

One simple but effective strategy to improve the accuracy of numeric estimates such as forecasts is to use the average of multiple estimates (Ashton 1986, Clemen 1989, Hogarth 1978), taking advantage of a statistical phenomenon that has been popularized as the "wisdom of crowds" (Almaatouq, Noriega-Campero, et al. 2020, Atanasov et al. 2017, Becker et al. 2017, Budescu and Chen 2014, Chen et al. 2014, Da and Huang 2020, Keuschnigg and Ganser 2016, Minson et al. 2018, Palley and Soll 2019). A central practical question is how communication between group members impacts the accuracy of the resulting "crowd" estimate (Atanasov et al. 2017, Becker et al. 2017, Chen et al. 2014, Da and Huang 2020).

Despite a common theoretical expectation that groups produce the most accurate estimates when people are independent (Budescu and Chen 2014, Hogarth 1978, Lorenz et al. 2011, Palley and Soll 2019, Surowiecki 2004), experimental research has found that communication can increase the accuracy of numeric estimates under carefully controlled conditions (Almaatouq, Noriega-Campero, et al. 2020, Atanasov et al. 2017, Becker et al. 2017, Jayles et al. 2017, Minson et al. 2018). While this body of research has provided strong evidence that social exchange can sometimes improve the wisdom of crowds, these experiments have produced contradictory results on what exactly is required for estimate accuracy to improve (Hastie 1986). How controlled must group interaction be in order for social learning to occur?

Recent laboratory experiments demonstrating benefits of communication followed a procedure where participants only exchange the numeric value of their estimates (Almaatouq, Noriega-Campero, et al. 2020, Becker et al. 2017, 2019, Jayles et al. 2017) and are comparable to early laboratory implementations of the "Delphi method" (Dalkey 1969, Gustafson et al. 1973, Ven and Delbecq 1974). However, requiring such strictly controlled communication limits the scope of practical applications, and implementation requires understanding whether these benefits can be obtained through informal discussion, e.g., in a committee meeting.

Despite the importance of determining whether collective intelligence requires carefully mediated communication or whether it can happen in everyday conversation, attempts to answer this question have yielded contradictory findings. Some findings suggest that numeric exchange is optimal while others suggest that informal discussion is optimal (Hastie 1986). We argue that this confusion results from the assumption that communication has a single main effect—i.e., that social influence is always either helpful or harmful; and that discussion is always either better or worse than numeric exchange.

By adopting a network model of opinion formation (Almaatouq, Rahimian, et al. 2020, Becker et al. 2017, DeGroot 1974, Golub and Jackson 2010) we are able to show how these apparent contradictions can be explained as expected outcomes for different points in a previously undescribed parameter space. This network model predicts that in decentralized networks where everyone is equally influential, communication is expected to reliably improve the accuracy of the average estimate (Becker et al. 2017, 2019, Madirolas and de Polavieja 2015). In contrast, centralized networks will improve only when the central node pulls the average estimate towards the true answer (Becker et al. 2017). The likelihood that the influence of central nodes will increase accuracy depends on the pre-discussion belief distribution.

Our key insight is that informal discussion acts like a centralized network, due to the availability of mechanisms which allow people to obtain disproportionate influence such as persuasion or talkativeness. When centrality—i.e., influence—is uncorrelated with accuracy, the relative improvement of centralized and decentralized networks is determined by the probability that any randomly selected individual would influence the group in the correct direction.

Based on this network theoretical model of social influence, we argue that the effect of discussion—whether it helps or harms accuracy—depends on the statistical properties of the pre-communication estimate distribution. In contrast, we expect from prior empirical and theoretical findings that communication limited to the exchange of numeric estimates will reliably increase the accuracy of the average estimate (Becker et al. 2017, 2019, Madirolas and de Polavieja 2015). However, conversation may under favorable conditions improve accuracy even more than numeric exchange. As a result, we expect that numeric exchange will outperform informal discussion for some estimate distributions, and vice versa for other distributions. In other words, the optimal communication format will depend on the estimation task itself.

This theoretical perspective thus provides sufficient conditions to explain contradictions in prior research on whether and when communication improves numeric estimate accuracy. This argument also highlights how simple practical interventions can be used to improve the quality of group decisions.

The rest of this paper proceeds as follows. First, we provide a review of network models of opinion formation and derive a simple heuristic to predict when discussion will help or harm numeric estimate accuracy. Then, we test our theoretical predictions empirically. We first test these predictions through a reanalysis of publicly available experimental data, though these prior experiments were not designed for the present analysis. We then support our argument with a pre-registered experiment designed to span a wide variety of estimation tasks producing a large variance in relevant statistical characteristics. We conclude with a more detailed discussion of practical implication and consider how a network analytic framework may guide future research and intervention design.

## 2. Theory and Hypotheses
### 2.1. Background

Our hypotheses derive from research (Almaatouq, Rahimian, et al. 2020, Becker et al. 2017) following the DeGroot (1974) model of opinion formation. This model assumes that each individual in a population starts with some initial estimate, i.e., a response to some numeric estimate task. Each individual can then observe the numeric estimate of some or all other members of the population. People then update their estimate as a weighted average, combining their own initial estimate with the estimates of their peers. This weighting is completely flexible, in that it allows someone to ignore (or be disconnected) from some peers by assigning that peer a weight of zero, ignore their peers all together, copy a single neighbor, or adopt the group average. After updating, individuals then observe the revised estimate of their peers and update again.

The weight that each individual places on each of their peers can be encoded in a directed weighted social network. This social network is an influence network, indicating who influences whom and how much. DeGroot shows that if this revision process is repeated indefinitely, the group will (under broad conditions) asymptotically converge on a weighted average of initial independent estimates, with each individual's estimate counting proportionately to their "centrality" score[1] in the influence network. The centrality for each individual reflects both the number of peers they have, i.e. how many people they influence, as well as how influential they are on each peer.

Because individual contribution can be measured by their centrality score, the dynamics for the group as a whole can be characterized by the distribution of centrality scores. One useful summary statistic is the "centralization" score, which can be measured with a Gini coefficient (Badham 2013, Freeman 1978). (Where centrality is a property of a single network member, centralization is a property of the network as a whole.) When centralization is high, the group is said to be "centralized" and influence is concentrated within relatively few individuals.

---

[1] Influence is proportional to degree centrality after a single update and eigenvector centrality after infinite updates.

In highly centralized networks, the effect of social influence will depend only on the accuracy of the central nodes, generating the 'wisdom of the few' rather than the wisdom of the crowd (Almaatouq, Rahimian, et al. 2020, Becker et al. 2017). This occurs because the group converges around the belief of the influential individuals. In contrast, "decentralized" networks where influence is evenly distributed would be expected to converge on the simple, unweighted mean of individual estimates. However, in laboratory settings where communication networks give everyone an equal number of peers—i.e., equal network centrality as measured by the observer— groups engaging in simple numeric exchange were drawn toward the estimate of accurate individuals (Becker et al. 2017).

This dynamic can be explained by noting that simply counting a person's number of network connections isn't a full reflection of their network centrality. Even once a binary communication network is established—in terms of mapping who talks to whom—there are two ways a person's centrality in the weighted influence network may further vary. First, a person's persuasiveness on peers (the weight that peers place on them) may vary through factors such as status or argumentation. Second, people can vary in the influence their own belief has own themselves (self-weight) e.g. through stubbornness. Relative stubbornness leads to relative influence because others revise in response to a stubborn individual's belief, who remains themself unchanged.

In the DeGroot model, self-weight is operationalized as weight placed on one's own belief relative to that of peers. In the network map, this is a "self-tie" or "loop" (influence from oneself to oneself) thus giving oneself greater incoming network weight. Using this model, stubbornness or other sources of self-weight (e.g. confidence) can be formally interpreted as centrality. And in Becker et al.'s (2017) experiment, subjects who were more accurate also made smaller revisions, i.e. appeared more stubborn. As a result, communication which were decentralized when measured by who-talks-to-whom ended up as weakly centralized networks in which the accurate individuals were the central individuals.

These arguments may superficially seem to suggest that centralized communication networks are simply unpredictable as compared with decentralized communication networks, since accuracy in centralized networks will determined entirely by the selection of the central node. However, Almaatouq et al. (2020) demonstrated that accuracy in centralized networks can be reliably characterized via the statistical properties of the initial estimate distribution. To explain this result intuitively, consider the conditions in Becker et al.'s (2017) experiment: subjects are randomly assigned to a location in a 'star' network with one central node observed by all their peers. Assume, consistent with empirical data, that the overall movement of the group mean is relatively small compared to the range of estimates. Then after social influence, the group mean will have moved a relatively small amount in the direction of the central node. Thus, the probability that the group improves depends only on the probability that the central node is on the same side of the mean as the true answer.

This analysis is an approximation: it is possible for the mean to "overshoot" the true answer if the central node is in the right direction but also wildly inaccurate. However, Becker et al.'s data suggest that this approximation effectively characterizes empirical processes. Below, we detail how this intuition can be used to derive a heuristic based on an approximation of the DeGroot (1974) model of opinion formation. We then use this demonstration to generate testable hypotheses relating discussion to numeric exchange that support an empirical analysis of laboratory data. By adopting a network theoretic model of the effect of communication on estimate accuracy, we can explain why informal discussion will sometimes improve accuracy more than numeric exchange, and why the opposite effect will sometimes occur.

**2.2. Theoretical Derivation**

Our theoretical argument and empirical analysis is based on the observation that a simple heuristic can predict the probability that group discussion can improve the accuracy of the average estimate. This heuristic is φ, or the proportion of individuals whose estimates fall on the same side of the mean answer as the true answer. The following section derives this heuristic.

Assume for simplicity (following Almaatouq et al., 2020) a network with one high influence individual (with estimate H) and N-1 equally low influence individuals (with mean estimate $\bar{L}$).

Assume also (again following Almaatouq et al.) some estimate updating process (e.g. discussion) such that the mean estimate after communication can be calculated as a weighted mean, $\mu_{post} = CH + (1-C)\bar{L}$ where $C$ represents the centrality[2] of the high influence individual, which in DeGroot (1974) would be a weight proportional to network centrality. Let $\mu_{pre}$ be the pre-communication estimate, and note that $\mu_{pre} = \frac{1}{N}H + \frac{N-1}{N}\bar{L}$. Thus $C = \frac{1}{N}$ means that everybody is equally influential, and $C = 1$ indicates that the group simply adopts the high influence individual's estimate. Assume that $C \geq \frac{1}{N}$.

Let $\theta$ be the true answer. Define "group error" as the distance between the mean estimate (pre- or post-communication) and the true answer, i.e. $|\mu - \theta|$ where $\theta$ is the true answer. Note that, as observed by Becker et al., (2017), there are two primary cases of interest—when H and $\theta$ are on the same side of $\mu_{pre}$, meaning that H is "in the direction of truth" relative to $\mu_{pre}$; and when H and $\theta$ are on opposite sides $\mu_{pre}$. When H is on the opposite side from truth, the group error will increase. When H is on the same side as truth, then group error will decrease, i.e. the group will become more accurate, as long as $\mu_{post} - \mu_{pre} < 2(\theta - \mu_{pre})$. Whether this condition holds depends on the value of H and C, which together determine $\mu_{post}$.

When $H - \mu_{pre} < 2(\theta - \mu_{pre})$, i.e. when the central individual is more accurate than the mean, then communication will increase group accuracy regardless of the value of C. In contrast, when $H - \mu_{pre} > 2(\theta - \mu_{pre})$, the central individual is in the direction of the truth but also very inaccurate relative to the group as a whole. In such a case, a highly influential H may cause the group to "overshoot" the true value so far that it initially moves towards the true value but then moves past the true value and the group ultimately ends up less accurate than it started. This will

---

[2] Almaatouq et al. (2020) present a slightly rearranged version of the weighted sum in their equation S.5: $\mu_{post} = \omega H + (1-\omega)(\frac{H}{N} + \bar{L})$. $\omega$ represents a form of network centralization, a measurement of the overall inequality in influence which we measure below as the Gini coefficient. Our measurement of network centrality, the influence of a single individual, is equivalent in this case as $C = \frac{1}{N} + \frac{N-1}{N}\omega$. We make this rearrangement of terms in order to represent the post-influence estimate as a weighted sum of H and L.

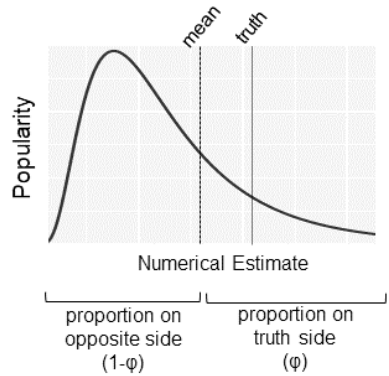

**Figure 1. Conceptual diagram showing the calculation for φ.**

*Notes: φ is the proportion of individuals on the truth side of the mean, a useful heuristic for predicting the effect of group discussion.*

occur for example when C=1, and the group simply adopts the central individual's estimate. More generally, there is some value $C'$ such that for all $C' \leq C \leq 1$ the group mean will be less accurate. Importantly, this also means that for all $C \leq C'$ the group will, despite being influenced by such a highly inaccurate individual, nonetheless improve by sheer chance.

As a critical step in deriving our heuristic, assume that the following is always the case: either H is on the opposite side of truth, and the group becomes less accurate; or H is on the same side of truth, and $C \leq C'$, and thus the group becomes more accurate. This assumption is consistent with experimental settings. Even where an individual is given highly centralized position in a communication network, they typically show relatively low C in the weighted influence network (e.g., people are influenced, but make relatively small revisions). In experiments with highly centralized "star" networks (with one highly centralized individual) groups did not completely adopt the estimate of the central individual, but rather only moved in the direction of that influential estimate (Becker et al., 2017).

Finally: Let φ indicate the proportion of individuals whose initial estimate is on the same side of the mean as the true answer, as shown in the conceptual diagram in Figure 1. Then, φ represents the probability that any randomly selected individual will hold an estimate on the truth side of the mean. When network centrality (influence) is uncorrelated with estimate (consistent with experimental data) then φ equals the probability that H is on the truth side of the mean and thus that communication in a centralized network will increase accuracy. This analysis yields a simple empirical heuristic. Centralized networks will improve in expectation when φ>0.5 and become less accurate when φ<0.5.

### 2.2.1. Empirical Hypotheses.

Using this framework, we can explain the dynamics of numeric exchange versus informal discussion. While numeric exchange does allow some variation in influence through stubbornness, (refusal to revise even as others do) this level of influence is inherently limited. In contrast, informal discussion allows people the opportunity to directly sway each other's estimates, allowing any individual potentially unlimited influence over group opinion. Thus, while both communication formats can be characterized as demonstrating emergent centralization (i.e. in otherwise decentralized communication networks), the possible extent of centralization is much greater for informal discussion. Informal discussion can, in theory, reach maximal centralization (one person dictating group estimates). As a result, the moderating effect of initial estimate distribution will be much stronger for informal discussion than for numeric exchange since this moderating effect is stronger in more centralized communication settings.

Taken together, these observations can explain the apparent contradiction in previous experimental findings. Assume that for some group, numeric exchange has a fixed probability ω of generating increased accuracy. If an experiment happens to be conducted with estimation tasks that yield initial estimate distributions with φ<ω, then informal discussion can be expected to decrease accuracy while numeric exchange increases accuracy; and vice versa when φ>ω. Previous studies have reported conflicting results on whether numeric exchange will reliably improve estimate accuracy. We therefore based our hypotheses on the assumption that ω≈1/2, in the absence of any alternative prior. Based on this theoretical perspective, we will test the following pre-registered hypotheses (see Appendix for registration details).

> **H1:** φ predicts probability of improvement for both discussion and numeric exchange.
> **H2a:** The effect of φ will be greater for discussion than for numeric exchange.
> **H2b:** Discussion will outperform numeric exchange when φ>1/2, and vice versa when φ<1/2.

## 3. Empirical Methods

We first tested these hypotheses on publicly available data from previous experiments (Becker et al. 2017, 2019, Gürçay et al. 2015, Lorenz et al. 2011). While one experiment tested informal discussion, and three experiments tested numeric exchange, no data available to us directly compared numeric exchange with discussion. As a result, this analysis compares both across experimental procedures as well as across question content, thus limiting the validity of causal inference in comparing the two types of communication. We therefore used this initial data to conduct statistical power tests and generate hypotheses for a pre-registered replication experiment to directly compare numeric exchange with informal communication. In this new experimental data, collected using the Empirica platform (Almaatouq et al., 2020), we also pre-tested tasks to ensure that our experimental trials included a wide range for $\varphi$. This method allows us to identify potential heterogeneity in the effect of social influence that may have been overlooked in previous research. Table 1 shows the distribution $\varphi$ for both reanalysis and replication trials.

### 3.1. Reanalysis

To conduct an initial test for the explanatory power of this model, we first reanalyzed data from previous experiments that measured estimate accuracy in groups before and after interaction. This analysis uses four previously published studies (Becker et al. 2017, 2019, Gürçay et al. 2015, Lorenz et al. 2011). These datasets were all made publicly available through the initial publications.

|  | Reanalysis | Replication |
|---|---|---|
| Toward ($\varphi>0.5$) | 63% | 49.5% |
| Away | 29% | 46.5% |
| Split | 8% | 4% |

**Table 1. The distribution of trials by the majority opinion as measured by φ.**

*Notes: Data is "split" when $\varphi=0.5$ While the data in previous experimental trials was heavily skewed toward a majority-correct task set, our replication experiment successfully produced a mixed balance of outcomes for φ.*

Detailed methods can be found in the initial publications, and each study follows a similar procedure. Subjects were asked to complete estimation tasks (e.g. visual estimation, trivia questions, and political facts) before and after exchanging information via a computer mediated communication process. An example of visual estimation task is an image of a jar of gumballs where subjects are asked to estimate how many gumballs are in the jar. An example of a trivia question is estimating the length of the border of Switzerland. An example of a political fact is asking subjects to estimate the number of undocumented immigrants living in the United States.

In three of the studies (Lorenz et al, 2011; Becker et al, 2017; Becker et al, 2019) subjects only exchanged numeric estimates. These studies therefore represent a method equivalent to a digitally mediated version of the Delphi method. Lorenz et al. (2011) allowed 5 rounds of revision (1 independent estimate and 4 socially influenced estimates) while Becker et al. (2017; 2019) allowed 3 rounds of revision (1 independent estimate and 2 socially influenced estimates). In contrast, Gürçay et al allowed subjects to engage in continuous, informal discussion via a computer chat interface, so that subjects provided only two answers, a pre-discussion and a post-discussion estimate (1 independent estimate and 1 socially influenced estimate). The data from Gürçay et al. is missing chat transcripts from 5 groups, and therefore those trials are omitted from analyses where the chat transcripts are necessary (i.e., measuring emergent centralization).

## 3.2. Replication Study

Our replication study follows the same general research paradigm as previous experiments, with two key design features. First, we pre-tested questions to ensure that our tasks covered a wide range of φ, in order to identify heterogeneity that may have been overlooked in previous research comparing informal discussion and numeric exchange. Second, our design allows for direct comparison between informal discussion and numeric (Delphi) exchange, where previous studies in our dataset only allowed either one or the other. Following previous research, subjects first provided an initial independent estimate for a numeric estimation task, then engaged in communication, and then provided a final estimate. Subjects were paid based on their accuracy. A complete list of questions is provided in the Appendix.

Subjects were recruited via Amazon Mechanical Turk and were randomly assigned either to informal discussion groups or numeric exchange groups. This method allowed us to run the two experiments in parallel while maintaining random assignment to conditions. Subjects were recruited in batches by sending an email with a link to access the experimental website. All subjects who arrived at the website at for a given experimental session were randomly assigned to a group of 20 individuals. If the number of subjects who arrived at the page was not divisible by 20 (e.g., if 59 subjects arrived) individuals would be randomly assigned either to participate or not. Subjects who were not assigned to a group were returned to the pool and invited at a later time.

For informal discussion, subjects were given 60 seconds to read the question and provide their initial estimate. They were then placed into a chatroom with the other subjects in their group and given 3 minutes to discuss the question. After this period, subjects were given 30 seconds to provide their final answer. At each stage, a countdown timer indicated the time remaining for that stage.

For numeric exchange, subjects were given 60 seconds to read the question and provide their initial estimate. Subjects were then shown a list of the answers provided by other subjects in their group as well as the average of answers provided by other subjects. The presented order of peer answers was randomized for each subject. Subjects were given 60 seconds to review the responses of other subjects and provide an updated (second) estimate. This process was then repeated twice, allowing subjects to observe the revised answers of their peers and provide a third and then fourth estimate in the same fashion. Finally, subjects were shown the revised answers of their peers and given 30 seconds to provide a fifth and final response. At each stage, a countdown timer indicated the time remaining for that stage.

By this method, subjects in each condition had 60 seconds to provide an initial estimate, 3 minutes to respond to the estimates of their peers, and 30 seconds to provide a final answer. Thus all subjects in total had 4.5 minutes between initially viewing the question and providing their final answer. In total, we collected 10 trials (i.e., 10 groups of 20 individuals) each for 10 unique questions for each condition, producing a total of 100 trials of informal discussion and 100 trials of numeric exchange. Each subject participated in only one question for only one condition. In total, we collected data from 4,000 unique subjects. This pre-registered sample size was based on power tests using the re-analysis of previous data to achieve 80% power on H2.

### 3.3. Analysis

For the purposes of our analysis, a single experimental trial consists of a single group of individuals completing a single estimation task. For each experimental trial, we ask a simple question: after social influence, was the mean estimate more accurate, i.e. closer to the true answer? Our primary outcome of interest, for any given experimental condition, is the proportion of trials in which the average answer became more accurate. We assess these outcomes using one- and two-sample proportion tests as well as logistic regression. For our pre-registered reanalysis of previously published data, we used cluster-robust logistic regression with fixed effects intercepts due to the structure of the data, since each recruited group answered multiple questions (i.e., completed multiple trials). For previously published data on numeric exchange, we combined all the data and analyzed it as if collected in a single experiment, clustering on dataset as well as groups within each dataset.

In order to test our theoretical mechanism, that the effect of $\varphi$ is moderated by centralization, which measures the relative concentration or equity of the distribution of network centrality. We note that there are many ways to measure centrality, including the popular metric of "degree centrality" which simply counts the number of connections a person has. However, in our present context of all-to-all communication (such as a committee discussion) this metric would indicate that our discussion networks are perfectly decentralized—everyone is equally connected. However, we wish to reflect the influence or weight an individual has on group opinions, not just their location in the communication network, and so we measure centrality for each individual in the discussion groups as the number of chat messages they sent.

For the group as a whole, we follow standard practice and measure network centralization as the Gini coefficient on centrality scores (Badham 2013). We note that the relationship between $\varphi$ and centralization was statistically uncertain in our reanalysis of prior data (see Appendix) and would require an infeasibly large sample to achieve statistical power, and our replication experiment is therefore powered only for our main hypotheses H1 and H2. The goal of this underpowered analysis is therefore to show that empirical data is consistent with theoretical expectations even if we cannot reject the null hypothesis.

The analysis presented here differs slightly from our pre-registered analysis in order to simplify the presentation of results, improve robustness, and report additional tests of interest. We report these corrections in the Appendix along with full details for the pre-registered analysis. We find comparable results for both our reanalysis of previous data as well as our replication analysis, and therefore present the results of both analyses simultaneously.

## 4. Results

The primary question facing many prior researchers, for any given communication format, is simply "does this process increase estimate accuracy?" We therefore begin our analysis by testing whether there is a main effect of social influence for each condition in each dataset. However, as expected, we do not find any consistent pattern. In our reanalysis, we found that social influence increased accuracy for a majority of trials in both numeric exchange (61% improvement, $P<0.001$, proportion test) and informal discussion (55% improvement, $P=0.11$, proportion test). In our replication, numeric exchange did not have any clear main effect on accuracy (45% improvement, $P=0.37$, proportion test) while discussion again increased accuracy (54% improvement, $P=0.48$, proportion test). These results are consistent with the contradictory results of previous literature, and consistent with our expectation that social influence does not have a single main effect on estimate accuracy. We note that while some recent research reports a consistent improvement in numeric exchange (Becker et al. 2017, 2019), this effect relies on a positive correlation between estimate error and stubbornness, i.e. response to social information (Becker et al. 2017, Madirolas

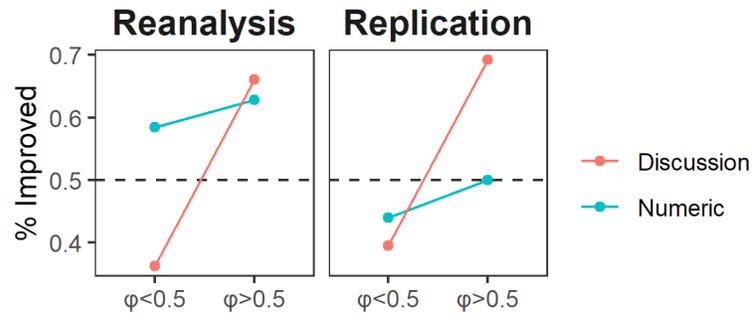

**Figure 2. The relative benefits of informal discussion vs. numeric exchange depend on whether φ<1/2.**

*Notes: The probability that informal discussion will improve accuracy depends on the majority opinion relative to the initial mean and the true value, i.e. whether or not φ<½. In contrast, φ has a minimal effect on numeric exchange, with results varying across datasets but not substantially varying within datasets.*

and de Polavieja 2015). Our experiment replicates this dynamic, i.e. groups improve when that correlation is negative and less accurate when that correlation is positive (Figure A1, Appendix).

We next test H1 ("φ predicts probability of improvement for both discussion and numeric exchange") i.e. the hypothesis that the effect of social influence is determined by the initial estimate distribution. To test this hypothesis, we estimate the effect of φ on the probability that social influence improves estimate accuracy. Our pre-registered analysis consists of a logistic regression predicting the binary outcome "did error decrease" as a function of φ. In both the re-analysis and the replication analysis, we found that the probability of improvement increases with φ for both discussion and numeric exchange ($P<=0.01$, all four tests, table A2). This analysis shows that the effect of social influence is determined by the initial estimate distribution but does not yet explain the inconsistent comparisons of numeric exchange and informal discussion.

### 4.1. Resolving Contradictions in Prior Findings

Consistent with our hypothesis that the effect of the pre-communication estimate distribution is magnified by network centralization, we find that the effect of φ is greater for informal discussion than for numeric exchange, supporting H2a ("The effect of φ will be greater for discussion than for numeric exchange.)

Figure 2 shows the probability of improvement as a function of the majority opinion by dividing outcomes based on whether $\varphi > \frac{1}{2}$ (majority correct) or $\varphi < \frac{1}{2}$ (majority incorrect). For informal discussion, the pre-communication estimate distribution determined whether information exchange helped or harmed accuracy. In discussion trials, the majority opinion significantly predicted outcomes (P<0.01 both datasets, proportion test): when $\varphi > \frac{1}{2}$, discussion increased accuracy in 66% of reanalysis trials and 69% of replication trials, but when $\varphi < \frac{1}{2}$ discussion increased accuracy in only 36% and 40% of trials, respectively.

In contrast, the majority opinion has a negligible effect on improvement for numeric exchange (P>0.44, both datasets, proportion test). While the reanalysis and replication data disagreed regarding the main effect of numeric exchange (suggesting no main effect at all), the location of the majority opinion did not tip the scales for numeric exchange in either sets of trials.

Most importantly, Figure 2 shows how previous experiments could at times yield conflicting results, supporting H2b ("discussion will outperform numeric exchange when φ>1/2, and vice versa when φ<1/2."). When the majority is away from truth, numeric exchange improves accuracy more than discussion (P<0.01, reanalysis; P=0.82, replication, proportion tests); when the majority is toward truth, discussion is superior (P=0.59, reanalysis; P=0.06, replication, proportion tests). While these two-sample tests are not statistically clear in every case, the overall effect is consistent across replication: when $\varphi > \frac{1}{2}$, the probability of improvement in discussion is greater than numeric exchange for all cases, and vice versa when $\varphi < \frac{1}{2}$.

**4.2. Testing Mechanisms: Centralization**

Our analysis thus far is sufficient to explain prior, seemingly contradicting results in which discussion is sometimes better and sometimes worse than numeric exchange. As a further analysis, we test whether results are consistent with our hypothesized mechanism. The preceding analysis was motivated by the theoretical expectation that emergent centralization in discussion networks amplifies the effects of pre-communication estimate distribution. To measure emergent centralization in informal discussion, we calculate the Gini coefficient on individual contribution as measured by the number of chat messages sent. We then test for an interaction between

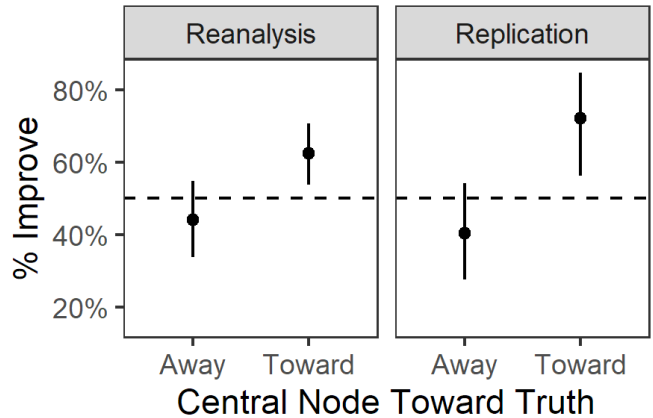

**Figure 3. Improvement as a result of central node belief.**

*Notes. Proportion of discussions where group accuracy improved depending on whether the pre-communication estimate of the most talkative (central) person holds a belief on the truth side of the group mean.*

centralization and φ in a logistic regression, finding results that are statistically uncertain but in the correct direction for both our reanalysis and our replication. While this interaction term was statistically uncertain in the data from previous experiments, it was consistent with theoretical expectations. We find comparable results in our replication dataset despite collecting fewer samples, supporting this theoretical model ($P=0.31$, $0.32$ respectively; see Table A3).

As a direct test for the effect of centralization, we estimate the extent to which the most talkative (central) person in each group predicts group outcomes, as shown in Figure 3. To measure this, we determine for each discussion trial who the most talkative person is, and whether their pre-discussion estimate is in the correct direction compared to the group mean. We then measure whether groups are more likely to improve when the most talkative (central) individual's pre-discussion estimate is in the direction of truth. We find for both reanalysis and replication that discussion is significantly more likely to improve group accuracy when the most talkative person is in the correct direction ($P<0.01$, both analyses, proportion test).

What these results show is that talkative people are indeed influential, i.e. central, in discussions. Because prior research has found a correlation between confidence and accuracy, we might also expect a correlation between talkativeness and accuracy. To explain why discussion does not reliably improve accuracy, but is instead determined by pre-discussion belief distribution, we test whether individuals who are more initially accurate are likely to be more talkative. To compare across tasks with different error ranges, we convert each person's error and talkativeness into quartiles[3] grouped by task (for error) and dataset (for talkativeness). We then measure the correlation between these two values (using cluster-robust regression as above for the re-analysis, clustering data by trial) finding a near-zero correlation between talkativeness and accuracy for both datasets (-0.01, $P>0.64$ for the reanalysis, 0.05, $P>0.06$ for our replication).

## 5. Discussion

Our results offer an explanation for why prior research sometimes showed that mediated numeric exchange improved group accuracy more than informal discussion, and sometimes showed the opposite. We found in both prior data and a novel pre-registered experiment that informal discussion sometimes increased and sometimes decreased accuracy depending on φ, the proportion of individuals on the truth side of the mean. In contrast, numeric exchange is only minimally (if at all) impacted by φ. As a result, informal discussion sometimes outperforms numeric exchange and sometimes does not. Thus, the relative benefits of discussion versus numeric exchange depend on the estimation task being considered.

An overly simple interpretation of the dynamics we observe in discussion would be to call the outcome a "majority effect," due to the tendency for the statistical majority to predict outcomes. However, while the location of the majority predicts outcomes in expectation across many trials, any given discussion is determined not by the majority but the most central individual. In this way, the dynamic we describe differs fundamentally from majority rules models as in e.g. group polarization theory (Myers and Lamm 1976, Sunstein 2006).

## Limitations

One limitation of our analysis is that some results are statistically uncertain. Results are relatively clear for our main hypothesis regarding the differential effect of φ on discussion versus numeric exchange. The dynamics of informal discussion show a strong effect of φ consistent across both our reanalysis and our replication, and numeric exchange varies in the main effect of social influence but nonetheless shows a consistent and relatively weak moderating effect of φ. However, our attempt to directly measure the hypothesized mechanism, emergent centrality, generated statistically uncertain results as expected from our underpowered analysis. Specifically, we did not see a clear interaction effect between φ and centralization as measured by the Gini coefficient on talkativeness. Nonetheless, the effects were in the expected direction and were consistent across both our reanalysis and our replication. Moreover, our additional analysis showed that the estimate of central (talkative) individuals are highly predictive of group outcomes, providing further support for our hypothesized mechanism.

---

[3] Due to the relatively small group size, bins smaller than quartiles were not possible for all trials.

Importantly, our observation that outcomes depend on φ, as shown in Figure 2, is at least minimally sufficient to explain contradictions in previous research. Additional research may be able to further identify the detailed dynamics of emergent centralization in informal discussion. One important consideration is that message volume is a noisy proxy measure for influence centrality, which is determined by more than just talkativeness, and future researchers may identify more precise methods for measuring emergent networks.

The biggest challenge facing our research is the limited ability to drawn generalizable conclusions outside of the laboratory. In particular, our reliance on networks of anonymous strangers makes it difficult to generalize to organizational settings where people interact within the context of previously established networks. One limitation in this respect is that we measured only one type of centrality, talkativeness, while influence networks in practice will be determined by mechanisms such as group norms and status relations. On the one hand, this limitation suggests that our results may have underestimated the true effects of emergent networks, as group dynamics such as status and norms may operate more strongly and deeply than factors affecting the interaction of anonymous strangers. As a result, we may reach a fairly reliable conclusion that informal discussion in organizational settings acts like centralized networks, and that the effect on accuracy is likely to vary widely by task. However, this limitation also means additional research is needed to understand the extent to which these dynamics are predictable and thus what may be optimal strategy in any given situation.

**Practical Implications and Future Research**

We hope that by clarifying the effects of communication format—discussion vs numeric exchange—this theoretical approach will help to draw a bridge between laboratory results on estimate accuracy and organizational communication in practice.

Our trials with numeric exchange produced a null result regarding a main effect on accuracy that neither falsifies prior claims that social influence improved accuracy nor supports them. Because this result may be explained by chance alone (as is the nature of statistical uncertainty) we refrain from speculating on what may have differed about our experimental setup. Broadly speaking, one possibility is that some features of our questions disrupted the correlation between confidence and accuracy; or that some higher-level effect disrupted the link between this mechanism and group dynamics, e.g. if people generally made less use of social information. Importantly to the present study, we find that the effect of numeric exchange for any given experimental procedure is relatively reliable across tasks, i.e. does not vary significantly as a result of the pre-exchange estimate distribution. Another possibility is that there is no main effect of numeric exchange, and the reliability of mechanisms identified in prior research varies by context. For those contexts where numeric exchange can be expected to improve accuracy (which could be determined empirically) it thus may represent a reliable technique for a group to aggregate opinions. In contrast, the benefits of discussion may vary from one estimation task to the next.

While our results reveal informal discussion to be unreliable, conversation also offers potential benefits beyond numeric exchange, if for example one group member possesses valuable information (Stasser and Titus 2003). By identifying the importance of emergent centralization and task properties—i.e. the value of a network model—our analysis paves the way for future research to more clearly identify the risks and benefits of group discussion, as compared to the inconsistent results of prior research. Our research also highlights a simple yet powerful intervention that may allow groups to harness the benefits of discussion while mitigating the risks: ensure equal participation by all contributing members.

A network theoretical framework highlights a possible communication strategy that can allow discussion while mitigating the risks of influential individuals. By embedding individuals in small interaction groups with just a handful of members, a group may maintain overall connectivity via overlapping group membership while nonetheless limiting the ability for any one individual to become overly central. This strategy also has the benefit of scaling to any size population, whereas all-to-all discussions such as a committee meeting are not feasible for large groups. By showing how network models can help explain the role of communication on belief accuracy, we hope that our research will enable the identification and testing of these and other possible interventions to optimize group estimate accuracy.

## 7. Appendix

Replication data and code available at:

https://github.com/joshua-a-becker/emergent-network-structure

Pre-registration available at:

https://osf.io/pvu9e/?view_only=2c63a8b7e6154139a0641f4e57f8f9f2

| Majority Relative to Truth | Format | Replication | % Improved | P.val |
|---|---|---|---|---|
| Away | Delphi | reanalysis | 0.584 | 0.171 |
| Away | Delphi | replication | 0.44 | 0.480 |
| Away | Discussion | reanalysis | 0.363 | 0.008 |
| Away | Discussion | replication | 0.395 | 0.222 |
| Toward | Delphi | reanalysis | 0.629 | 0.000 |
| Toward | Delphi | replication | 0.489 | 1.000 |
| Toward | Discussion | reanalysis | 0.661 | 0.000 |
| Toward | Discussion | replication | 0.692 | 0.008 |

**Table A1**. Pre-registered proportion tests. Each row represents a one-sample proportion test for a different part of the parameter space.

**Pre-registration and Analysis**

The pre-registered tests for the finding shown in Figure 2 included a series of descriptive one-sample and two-sample proportion tests for the probability of improvement for each point in the figure. However, to simplify the readability of our manuscript, we focus the main text on more central tests and include only the two-sample proportion tests for each line in Figure 2. Here we present the full pre-registered descriptive analysis. Table A1 presents a series of one-sample proportion tests, asking at each point in the parameter space whether social influence significantly improves (or decreases) group error.

The main text also reports a slightly modified version of our test for an interaction effect between φ and talkativeness centralization. Our pre-registered analysis only calculated the Gini coefficient for participants present in the discussion, which incorrectly omitted zeros from the calculation—i.e., people who did not contribute to discussion (and thus had minimal influence) but nonetheless contributed to the collective estimate by including an opinion. We note that while this test was statistically significant in its original version, it is not significant in the revised version. We provide code for both tests in our replication materials.

| | Outcome: Mean Closer to Truth (Yes/No) | | | |
|---|---|---|---|---|
| | Replication | | Reanalysis | |
| | Disc. | Delphi | Disc. | Delphi |
| Φ | 4.08** | 3.13* | 4.86*** | 5.60*** |
| | (1.38) | (1.22) | (0.96) | (0.51) |
| AIC | 122.90 | 138.43 | 400.12 | 431.59 |
| BIC | 151.55 | 167.09 | 478.52 | 749.83 |
| Log Likelihood | -50.45 | -58.22 | -179.06 | -129.80 |
| Deviance | 100.90 | 116.43 | 358.12 | 259.59 |
| Num. obs. | 100 | 100 | 309 | 299 |

***$p < 0.001$; **$p < 0.01$; *$p < 0.02$

**Table A2. Regression results showing effect of phi on probability that the error of the mean estimate (absolute deviation) decreases.**

*Notes: For the result with \*, P=0.01 (P<0.02)*

**Supplemental Analysis**

Table A2 shows the regression results for the effect of Phi. Table A3 shows the regression results for an interaction between phi and centralization as measured by Gini coefficient. Figure A1 compares the dynamics of numeric exchange across all four datasets by measuring the stubbornness/error correlation.

**Replication Question Text**

**Crowdfunding 1.** Consider this crowdfunding campaign: The goal of this app is to promote new music discovery in a fun and different way. This app would allow musicians to "drop" songs at specific physical locations. Anyone using the app would then be able to listen to the song by visiting that location. The app sought £30,000 (British pounds) and offered funders equity in the company, with a total equity of 35% for the whole campaign. How much money do you think the campaign raised? Answer: 30,000. Source: https://www.seedrs.com/gigdropper

**Crowdfunding 2.** Consider this crowdfunding campaign: The product is headphones designed for dance music. The goal of the product is to replicate the sound style of being in a club or party. The campaign followed a successful prior round of funding, and the company has already sold thousands of units. This campaign sought an additional £100,000 (British pounds) in exchange for

|  | PreReg. | | Revised | |
|---|---|---|---|---|
|  | Rean. | Rep. | Rean. | Rep. |
| Φ | -7.16 | -2.99 | 0.54 | -2.99 |
|  | (4.04) | (8.89) | (3.68) | (8.89) |
| Gini | -20.03** | -10.95 | -6.07 | -10.95 |
|  | (7.39) | (15.13) | (4.55) | (15.13) |
| Gini * Φ | 34.68** | 21.67 | 8.07 | 21.67 |
|  | (12.32) | (27.20) | (8.00) | (27.20) |
| AIC | 285.42 | 126.22 | 292.69 | 126.22 |
| BIC | 350.66 | 160.09 | 357.93 | 160.09 |
| Log Likelihood | -123.71 | -50.11 | -127.34 | -50.11 |
| Deviance | 247.42 | 100.22 | 254.69 | 100.22 |
| Num. obs. | 229 | 100 | 229 | 100 |

Outcome: Mean Closer to Truth (Yes/No)

***$p < 0.001$; **$p < 0.01$; *$p < 0.05$

**Table A3.** Regression results showing interaction effect between phi and centralization (Gini coefficient) on probability that the error of the mean estimate (absolute deviation) decreases. Only 229 datapoints are used from the prior experimental data due to missing chat data.

equity in the company, and ended up exceeding their goals. How much money do you think the campaign raised? Answer: 142,770. Source: https://www.seedrs.com/pump-audio

**Socio-Economic 1:** In 2009, approximately 690 million passengers boarded a plane. (So a round-trip flight counts for 2 passengers boarding.) How many of these passengers boarded out of an airport in the New York City area (JFK, La Guardia, and Newark)? (Give your answer in millions—e.g., enter 1 for 1 million.) Answer: 41. Source: http://infochimps.org/datasets/d35-million-us-domestic-flights-from-1990-to-2009

**Socio-Economic 2:** Across all colleges where the US Department of Education collected data, the average tuition revenue per full time (or equivalent) student was $10,438 per year. In terms of dollars, how much money do you think was spent on instruction, per student? Answer: 7912. Source: https://collegescorecard.ed.gov/data/

**Art 1:** [Image of Planteuse des Betteraves] This drawing by Vincent Van Gogh sold at auction in May, 2018. It is 18 inches tall by 20 inches wide, charcoal on paper. How much did it sell for? (Answer in millions of dollars, e.g. enter 1 for $1 million or 0.5 for $500,000) Answer: $3.6

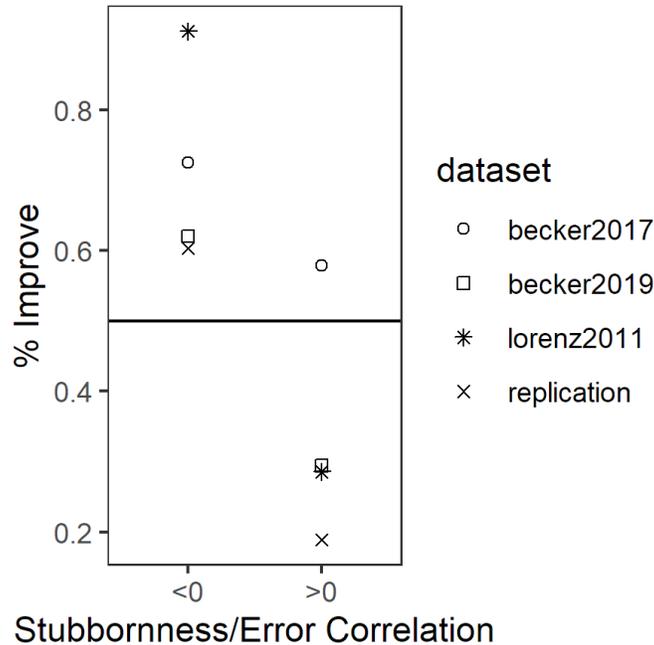

**Figure A1. Effect of Stubbornness Centralization on Improvement in Numeric Exchange**

*Notes. When the stubbornness/error correlation is less than zero, people with high error are likely to have low stubbornness, i.e. revise and thus the group improves.*

million. Source: https://www.christies.com/lotfinder/Lot/vincent-van-gogh-1853-1890-planteuse-de-betteraves-6134215-details.aspx

**Art 2:** [Image of La Lampe] This painting by Pablo Picasso sold at auction in November, 2018. It is 64 inches tall by 51 inches wide, oil on canvas. How much did it sell for? (Answer in millions of dollars, e.g. enter 1 for $1 million or 0.5 for $500,000). Answer: $29.6 million. Source: https://www.christies.com/lotfinder/Lot/pablo-picasso-1881-1973-la-lampe-6169488-details.aspx

**Geopolitics 1:** The Armed Conflict Location & Event Data Project (ACLED) is a non-governmental organization that tracks violent conflict in Asia, the Middle East, and Africa. One type of event they track is those where civilians were intentionally targeted. In 2018, they recorded 841 such events in Somalia. How many of this type of event do you think they recorded in Yemen for 2018? Answer: 609. Source: https://www.acleddata.com/data/

**Geopolitics 2:** The Armed Conflict Location & Event Data Project (ACLED) is a non-governmental organization that tracks violent conflict in Asia, the Middle East, and Africa. One type of event they track is those where civilians were intentionally targeted. In 2018, they recorded 841 such events in Somalia. How many of this type of event do you think they recorded in Syria for 2018? Answer: 1501. Source: https://www.acleddata.com/data/

**Gun Violence 1:** Gun Violence Archive (GVA) is a not for profit corporation that tracks gun-related violence in the United States. In 2018, GVA recorded 1,113 events in Baltimore. (A single event might involve more than 1 person.) How many events do you think they recorded in Chicago in 2018? Answer: 2812. Source: https://github.com/awesomedata/awesome-public-datasets#socialsciences

**Gun Violence 2:** Gun Violence Archive (GVA) is a not for profit corporation that tracks gun-related violence in the United States. In 2018, GVA recorded 1,113 events in Baltimore. (A single event might involve more than 1 person.) How many do you think they recorded in Philadelphia in 2018? Answer: 570. Source: https://github.com/awesomedata/awesome-public-datasets #socialsciences